\documentclass[12pt]{iopart}
\usepackage{graphicx}
\usepackage{epstopdf} 
\usepackage{dcolumn}
\usepackage{bm}
\usepackage{wasysym}
\usepackage{amssymb}
\usepackage{color}

\begin{document}
\bibliographystyle{unsrt}

\title[Bicoherence at the Low to High Confinement Transition in the TJ-II Stellarator]{Spatiotemporal and Wavenumber Resolved Bicoherence at the Low to High Confinement Transition in the TJ-II Stellarator}

\author{B.Ph.~van Milligen$^1$, T. Estrada$^1$, C. Hidalgo$^1$, T. Happel$^2$, E. Ascas\'ibar$^1$}
 \address{$^1$Asociaci{\'o}n EURATOM-CIEMAT para Fusi{\'o}n, Avda.~Complutense 40, 28040 Madrid, Spain}
 \address{$^2$Max Planck Institut f\"ur Plasmaphysik, Association Euratom-IPP, 85748 Garching, Germany}
\date{\today}

\begin{abstract}
Plasma turbulence is studied using Doppler reflectometry at the TJ-II stellarator.
By scanning the tilt angle of the probing beam, different values of the perpendicular wave numbers are probed at the reflection layer.
In this way, the interaction between zonal flows and turbulence is reported with (a) spatial, (b) temporal, and (c) wavenumber resolution for the first time in any magnetic confinement fusion device.

We report measurements of the bicoherence across the Low to High (L--H) confinement transition at TJ-II. 
We examine both fast transitions and slow transitions characterized by an intermediate (I) phase.
The bicoherence, understood to reflect the non-linear coupling between the perpendicular velocity (zonal flow) and turbulence amplitude,
is significantly enhanced in a time window of several tens of ms around the time of the L--H transition.
It is found to peak at a specific radial position (slightly inward from the radial electric field shear layer in H mode), and is associated with a specific perpendicular wave number ($k_\perp \simeq 6-12$ cm$^{-1}$, {$k_\perp \rho_s \simeq 0.8-2$}).
In all cases, the bicoherence is due to the interaction between high frequencies { ($\simeq 1$ MHz)} and a rather low frequency {($\lesssim 50$ kHz)}, as expected for a zonal flow.
\end{abstract}

\maketitle

\section{Introduction}
The spontaneous transition to the High confinement regime in fusion plasmas is a topic of great interest, due to its potential advantages for the design of a fusion reactor and the complex nonlinear physics involved.
At the TJ-II stellarator, the transition from Low to High confinement (L--H) is ``soft'' in the sense that the confinement improvement factor is small, and yet it possesses the typical features of any L--H transition (rapid drop of $H_\alpha$ emission, reduction of fluctuation amplitudes, formation of a sheared flow layer, ELM-like bursts)~\cite{Sanchez:2009,Estrada:2009}.

A common explanation for the L--H transition involves the generation of a zonal flow in a narrow radial region of the plasma near its edge. 
The zonal flow would be driven by turbulence via the Reynolds Stress mechanism in a strong pressure gradient region.
The growth of the (fluctuating) zonal flow then leads to the suppression of turbulence, creating a transport barrier, and the subsequent establishment of a steady state sheared flow with associated radial electric field that maintains the barrier~\cite{Burrell:1997,Diamond:2005,Wagner:2007}.
The detailed observation of this sequence of events is a challenge.

Due to the temporal and spatial scales involved in the L--H transition physics, specific experimental techniques are required to investigate the turbulence and flow dynamics~\cite{Estrada:2009}. 
At TJ-II, a two-channel Doppler reflectometer is available, allowing the simultaneous measurement of the perpendicular plasma velocity and density fluctuations at two radial positions, with good spatial and temporal resolution. 
In addition, the Doppler reflectometer offers the possibility to probe different turbulence scales (wave vectors) by steering the probing beam~\cite{Happel:2011,Estrada:2012b}.

\section{Methods} 
The TJ-II vacuum magnetic geometry is completely determined by the currents flowing in four external coil sets. 
At TJ-II, the normalized pressure $\left < \beta \right >$ is generally low, even in discharges with Neutral Beam Injection (NBI), and currents flowing inside the plasma are generally quite small (unless explicitly driven), so that the actual magnetic configuration and the rotational transform $\iota/2\pi$ are typically rather close to the vacuum magnetic configuration~\cite{Milligen:2011b}.
Here we focus on two magnetic configurations having $\iota(a)/2\pi=1.630$ and $1.553$.

The experiments have been carried out in pure NBI heated plasmas (line averaged plasma density $\langle n_e \rangle = 2-4 \times 10^{19}$ m$^{-3}$, central electron temperature $T_e = 300-400$ eV, {$T_i \simeq 140$ eV}). The NBI input heating power is kept constant at about $500$ kW during the discharge but the fraction of NBI absorbed power -- taking into account shine through, CX and ion losses, as estimated using a simulation code -- increases from 55 to 70\% as the plasma density rises. 

In this work, we consider two different L--H transition scenarios. 
Standard or {\it fast} L--H transitions are observed in the magnetic configuration having $\iota(a)/2\pi=1.630$~\cite{Happel:2011}, while {\it slow} transitions are observed in the configuration having $\iota(a)/2\pi=1.553$ and a low order rational (3/2) at $\rho = r/a \simeq 0.72$. 
In the latter configuration, the so-called intermediate phase (I), characterized by a predator-prey type interaction between turbulence and flows, appears between the L and H phases~\cite{Colchin:2002,Estrada:2010c,Estrada:2011}, which has also been seen in models~\cite{Kim:2003,Miki:2012} and on other devices~\cite{Conway:2011,Schmitz:2012}. 
In both scenarios, spatiotemporal and scale resolved Doppler reflectometry measurements were performed in series of repetitive discharges~\cite{Happel:2011,Estrada:2012b}. 
  
\subsection{Doppler reflectometry}
In Doppler reflectometry, a finite tilt angle is purposely introduced between the incident probing beam and the normal to the reflecting cut-off layer, and the Bragg back-scattered signal is measured~\cite{Hirsch:2001}. 
The amplitude of the recorded signal, $A$, is a measure of the intensity of the density fluctuations, $\tilde n$. 
By scanning the tilt angle of the probing beam, different perpendicular wave numbers are probed, thus allowing scale-resolved turbulence measurements. 
Furthermore, as the plasma rotates in the reflecting plane (flux surface), the scattered signal experiences a Doppler shift.
The size of this shift is directly proportional to the rotation velocity of the plasma turbulence perpendicular to the magnetic field lines, $v_\perp$, and therefore to the plasma background $E \times B$ velocity, provided the latter dominates over the phase velocity of density fluctuations (cf.~\cite{Estrada:2009}). 
The Doppler reflectometer signals, sampled at 10 MHz, allow determining $\tilde n$ and $v_\perp$ with high temporal and spatial resolution.

The relation between zonal flow growth, involving an interaction between high frequencies associated with drift wave turbulence and low frequencies associated with the zonal flow, and the cross bispectrum was elucidated in~\cite{Diamond:2000}.
Probing zonal flow dynamics therefore requires measuring the cross bispectrum between the radial and poloidal $E\times B$ velocities of high-frequency drift waves and the low-frequency zonal flow. 
Early attempts at observing the bicoherence across the L--H transition are found in Refs.~\cite{Milligen:1995b,Moyer:2001}.
As such measurements are quite difficult to perform in the plasma interior, the bicoherence is often calculated using density fluctuations only, which complicates the interpretation of the results. 
However, Doppler reflectometry allows the measurement of the turbulence fluctuation amplitude $\tilde n$ as well as the fluctuating perpendicular flow $\tilde v_\perp$, both with good temporal and spatial resolution, making the two main quantities involved in zonal flow dynamics accessible experimentally. 
In this work, we consider the complex amplitude Doppler reflectometry signal, $Ae^{i\phi}=(A \cos \phi , A \sin \phi$), which contains information of both the perpendicular flow, $\tilde v_\perp=d \tilde\phi /dt$, and the density turbulence, $\tilde n \propto \tilde A$. 

\subsection{Bicoherence analysis} 
The cross bicoherence of two signals $f(t)$ and $g(t)$ is defined as~\cite{Kim:1978}
\begin{eqnarray}
b^2_{fg}(\omega_1,\omega_2) = \frac{\langle |\hat f (\omega_1)\hat f (\omega_2)\hat g^* (\omega_1+\omega_2) | ^2\rangle}
{\langle |\hat f (\omega_1)\hat f (\omega_2) |^2\rangle \langle |\hat g (\omega_1+\omega_2) |^2 \rangle}, \nonumber
\end{eqnarray}
where $\hat f(\omega)$ is the Fourier transform of $f(t)$, a star indicates the complex conjugate, and the angular brackets refer to an average over equivalent realizations. In practice, the equivalent realizations are successive time sections, assuming the system is in steady state. The value of $b^2$ is bounded between 0 and 1, so that a high value indicates wave-wave coupling between the two base frequencies $\omega_1$ and $\omega_2$ and their sum frequency, $\omega_1 + \omega_2$. 
The auto-bicoherence is $b^2_f = b^2_{ff}$.

We also report the ``summed bicoherence'', defined as
$b^2_{\rm sum}(\omega) = \langle {b^2(\omega_1,\omega_2)} \rangle_{\omega_1+\omega_2=\omega}$,
and the ``total bicoherence'', defined as
$b^2_{\rm total} = \langle b^2(\omega_1,\omega_2) \rangle_{\omega_1,\omega_2}$.
To judge the relevance of a bicoherence value, it is compared to the expectation value of the bicoherence under the assumption that the signal is uncorrelated random noise.

\begin{figure}\centering
  \includegraphics[trim=0 0 0 0,clip=,width=8cm]{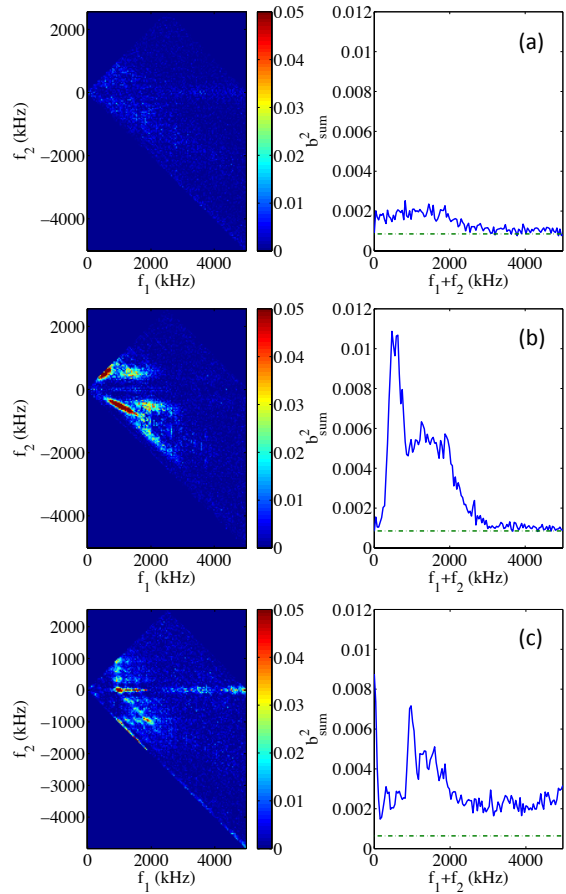}
\caption{\label{22277}Discharge 22277. Configuration with $\iota(a)/2\pi=1.704$. 
{Analysis of Doppler reflectometry signal.
(a) Mean auto-bicoherence and mean summed auto-bicoherence of complex Doppler signal in L mode ($1050 < t < 1080$ ms).
(b) Mean auto-bicoherence and mean summed auto-bicoherence of complex Doppler signal during the period of maximum bicoherence around the L--H transition ($1100 < t < 1130$ ms, $t_{L-H}=1116$ ms).
(c) Cross bicoherence between $\tilde v_\perp$ and $\tilde n$ in the same time window of (b).}
The dashed line indicates the noise level.}
\end{figure}

Computing the auto-bicoherence for the complex Doppler reflectometer data around the time of the L--H transition, one typically obtains results such as those shown in Fig.~\ref{22277}.
{Here and in the following, we use a frequency resolution of 128 (i.e., data records of 256 points to compute a single spectrum).
Fig.~\ref{22277}a shows the auto-bicoherence in the L phase, prior to the transition. The bicoherence level is low, close to the statistical noise level.
Fig.~\ref{22277}b shows the auto-bicoherence in a time window during the period of maximum bicoherence around the L--H transition.
It is very significant; its value is more than an order of magnitude above the noise level.}
The auto-bicoherence is concentrated along the line $f_2 \simeq -f_1/2$ (the `first harmonics' line), $f_2 \simeq f_1$ (interaction with a very small difference frequency), and also at $f_2 \simeq 0$.
These are indications that high  frequencies $f_a, f_b$ of the order of 1--2 MHz are interacting via a very small difference frequency $f_c = f_a-f_b$ (or harmonics $f_c = 2f_a-f_b$) such that $|f_c| \ll |f_a|, |f_b|$.
Frequencies above 1 MHz are not commonly observed in this context due to technical limitations; however, note that similar results were obtained at DIII-D~\cite{Moyer:2001}.

To facilitate the interpretation of the auto-bicoherence, {Fig.~\ref{22277}c} shows the cross bicoherence between $\tilde v_\perp$ and $\tilde n$ for the same discharge. 
Here, significant coupling is limited mainly to the lines $f_2 \simeq 0$ and $f_2 \simeq \pm f_1$, indicating an 
interaction between two very similar frequencies $f_a, f_b$, interacting via a very small difference frequency $f_c = f_a-f_b$ such that $|f_c| \ll |f_a|, |f_b|$, while harmonics are weak.
This is in fact what one would expect for the interaction between a zonal flow and turbulence~\cite{Diamond:2000,Nagashima:2006}.
The small frequency of the triplet is at or below the { lower} frequency resolution { limit} (50 kHz), whereas the two high frequencies are in the range from 1 to 2 MHz.
The fact that the cross bicoherence graph shows linear structures { and few harmonics}, while the auto-bicoherence {graph of the complex Doppler reflectometer data displays a strong response along the `first harmonics' line}, is a clear indication that the physical interaction primarily involves $\tilde v_\perp$ and $\tilde n$. 
Although the cross bicoherence between $\tilde v_\perp$ and $\tilde n$ is easier to interpret, 
the auto-bicoherence of the complex Doppler reflectometry signal (containing the same information) is easier to compute, 
not affected by the numerical errors inherent in the velocity calculation, and 
yields a larger signal to noise ratio in the total bicoherence value. 
Therefore, we will use the latter as a generic proxy to study the spatiotemporal evolution of the total bicoherence.


\section{Results}
\subsection{The fast L--H transition}
\begin{figure}\centering
  \includegraphics[trim=0 0 0 0,clip=,width=9cm]{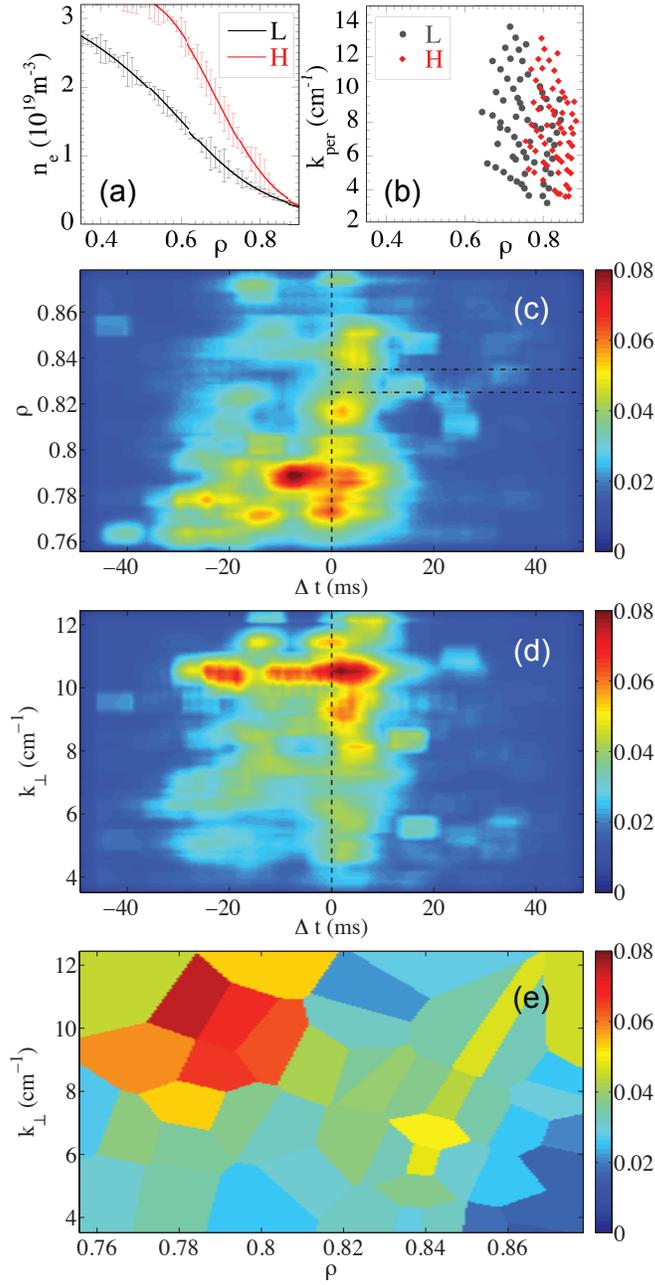}
\caption{\label{serie_101_42}Series of 25 discharges in the configuration with $\iota(a)/2\pi=1.630$. 
Total auto-bicoherence of complex Doppler reflectometry data. 
(a) Density profiles in the L and H phases.
(b) Position $\rho$ and $k_\perp$ of the measurement channels in the L and H phases. 
(c) Mean auto-bicoherence vs.~time and normalized radius, $\rho$.
(d) Mean auto-bicoherence vs.~time and $k_\perp$. 
(e) Mean auto-bicoherence vs.~$\rho$ and $k_\perp$ (averaged over $-10 \le \Delta t \le 10$ ms).
Radii and $k_\perp$ values of Figs. (c) -- (e) correspond to the H phase.
The vertical dashed line indicates the L--H transition; the horizontal dashed lines indicate the approximate position of the $E_r$ shear layer in the H phase. The statistical error level is about 0.006.}
\end{figure}

To study the radial and temporal behavior of the bicoherence across the fast L--H transition, we have analyzed a series of 25 discharges in the same magnetic configuration ($\iota(a)/2\pi=1.630$). See Ref.~\cite{Happel:2011} for more details about this series of discharges.
The Doppler reflectometer probing frequency and tilt angle were varied on a shot to shot basis, and correspondingly the radial position of the reflecting layer and the wave number $k_\perp$.
Fig.~\ref{serie_101_42}a indicates the evolution of the density profile across the L--H transition.
Due to the evolving profile, both the position of the reflecting layer and the wavenumber change as a function of time across the transition~\cite{Happel:2011}.
The measured channel locations and wave numbers are shown in Fig.~\ref{serie_101_42}b. 
Note that $k_\perp$ varies by less than 5\% between the L and H phases (matching black and red dots).
The $\rho$ and $k_\perp$ values indicated on the axes of Figs.~\ref{serie_101_42}c--e correspond to the situation in H-mode. 
Note that $\rho$ is a vacuum flux surface label; the actual confinement radius is somewhat less than $\rho=1$~\cite{Ascasibar:2005}.
The temporal resolution of the bicoherence calculation is 2 ms{, implying that each individual bicoherence estimate involves 78 independent spectral estimates}.

Figs.~\ref{serie_101_42}c--e show the values of the auto-bicoherence $b^2(k_\perp,\rho,\Delta t)$, averaged over $k_\perp$, $\rho$, and $\Delta t$, respectively.
It is seen that the bicoherence appears $\sim 20-30$ ms before the L--H transition at a specific radius ($\rho \lesssim 0.8$) and that specific perpendicular wave numbers are involved ($k_\perp \simeq 8-11$ cm$^{-1}$). 
The bicoherence persists until about $10-20$ ms after the L--H transition.
Similar behavior is observed across a wide range of magnetic configurations, showing that the phenomena are not associated with any specific rational surface.

It should be noted that turbulence is reduced at all { measured} scales, but most in intermediate scales, $k_\perp \simeq 7-11$ cm$^{-1}$, as reported in earlier work~\cite{Happel:2011}. After the transition, the radial electric field shear layer is located at $\rho \sim 0.83$, somewhat outward from the position of the maximum observed bicoherence values.

\subsection{The slow L--{I} transition}
\begin{figure}\centering
  \includegraphics[trim=0 0 0 0,clip=,width=9cm]{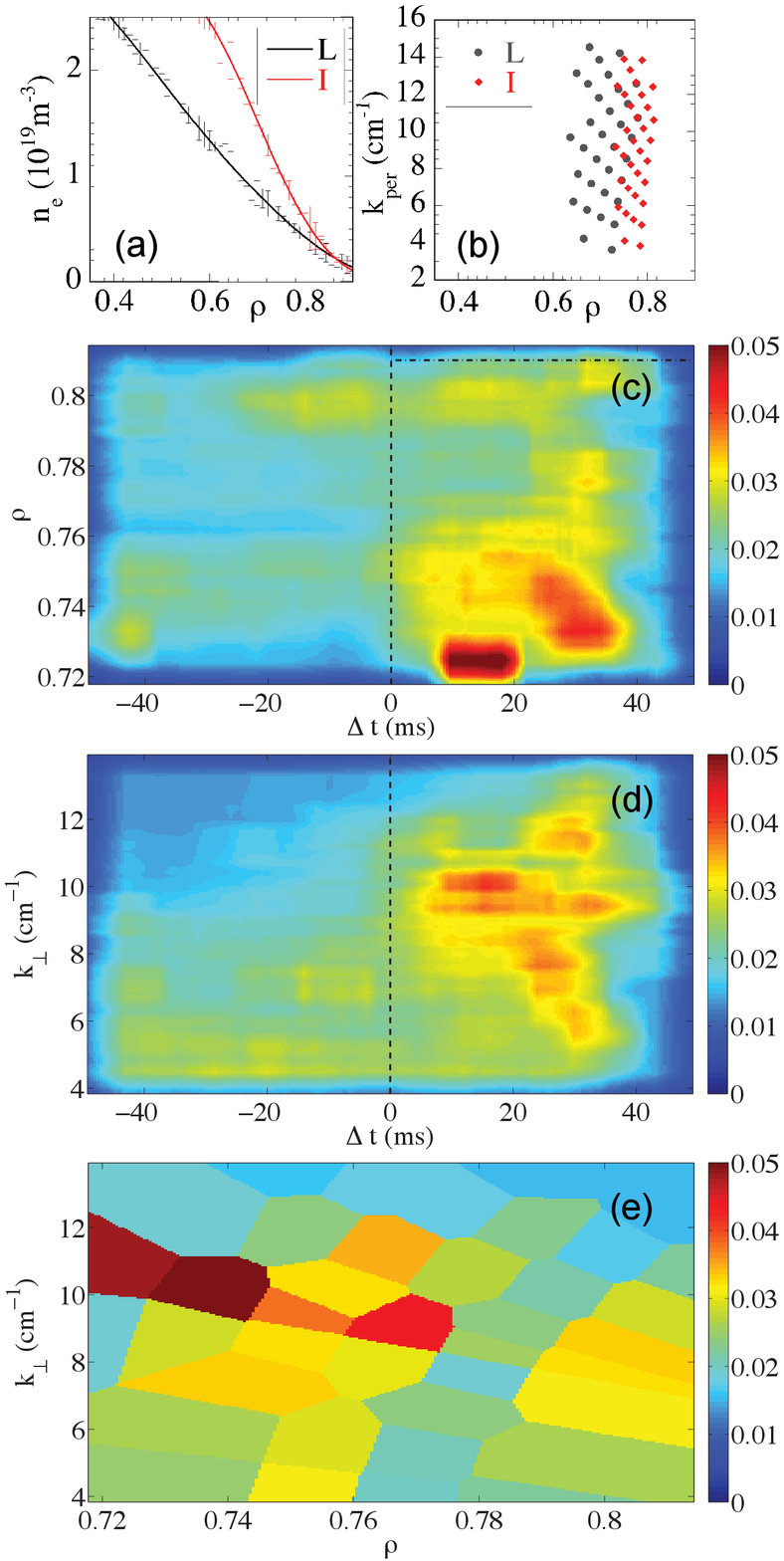}
\caption{\label{serie_100_35}Series of 26 discharges in the configuration with $\iota(a)/2\pi=1.553$. 
Total auto-bicoherence of complex Doppler reflectometry data. 
(a) Density profiles in the L and I phases.
(b) Position $\rho$ and $k_\perp$ of the measurement channels in the L and I phases. 
(c) Mean auto-bicoherence vs.~time and normalized radius, $\rho$.
(d) Mean auto-bicoherence vs.~time and $k_\perp$. 
(e) Mean auto-bicoherence vs.~$\rho$ and $k_\perp$ (averaged over $0 \le \Delta t \le 20$ ms).
Radii and $k_\perp$ values of Figs. (c) -- (e) correspond to the I phase.
The vertical dashed line indicates the L--I transition; the horizontal dashed line indicates the approximate position of the $E_r$ shear layer in the I phase. 
The statistical error level is about 0.006.}
\end{figure}

Fig.~\ref{serie_100_35} shows the auto-bicoherence for a different series of discharges in the configuration with $\iota(a)/2\pi=1.553$.
The elaboration of this figure is equivalent to that of Fig.~\ref{serie_101_42}.
The mean line average density at the transition is about 50\% lower than with the previous configuration.
Another difference with the previous result is that in these discharges, the transition at $\Delta t=0$ is not into the H-mode, but rather
into the intermediate (I) phase between the L-mode ($\Delta t < 0$) and the H-mode (starting much later).
We conjecture that this situation is due to a weaker zonal flow, incapable of producing the final transition into the H phase.
This would be consistent with the observation of the predator-prey relation between turbulence and flows, 
associated with wave numbers in the range $k_\perp \simeq 6-12$ cm$^{-1}$,
as confirmed by spectral analysis~\cite{Estrada:2012b}. 

The location of the auto-bicoherence peak appears to be somewhat further inward than in the previous section ($\rho < 0.76$).

\section{Discussion and conclusions}
In this work, we have used Doppler reflectometry measurements to show how the bicoherence evolves across the L--H transition at TJ-II.
It is the first time that turbulent fluctuation amplitudes and perpendicular (zonal) flow oscillations and their non-linear interactions have been observed simultaneously in the region of the plasma where the transport barrier of the L--H transition forms.  
The observation of the bicoherence does not appear to depend very sensitively on the magnetic configuration. 
A surprisingly clear picture emerges that is consistent with established ideas about the interaction between zonal flows and turbulence~\cite{Fujisawa:2007} in the important formation phase of the H-mode transport barrier.

In the case of the fast L--H transition, we have shown that the auto-bicoherence of the Doppler reflectometer signal is significant in a time window with a length of several tens of ms around the L--H transition time~\cite{Moyer:2001}, at a particular radial position ($\rho \lesssim 0.8$, inward from the radial electric field shear layer in H mode), and that a specific range of perpendicular wave numbers is involved ($k_\perp \simeq 8-11$ cm$^{-1}$). Here, $\rho_s = \sqrt{2m_iT_e}/eB = 0.1-0.2$ cm, so that this range corresponds to $k_\perp \rho_s = 0.8-2$.

A similar analysis in another set of discharges corresponding to a different magnetic configuration, in which the transition was characterized by an intermediate phase separating the L and H phases, the bicoherence appeared only after the L--I transition was made, and lasted several tens of ms afterward into the I phase. The perpendicular wave numbers involved ($k_\perp \simeq 6-12$ cm$^{-1}$) were very similar to the previous case, while the radial location of the bicoherence peak was somewhat further inward ($\rho < 0.76$).

In both cases, the auto-bicoherence was due to high frequencies (of the order of 1--2 MHz) interacting with rather low frequencies ($< 50$ kHz), as expected for a Zonal Flow (ZF)~\cite{Miki:2012}. With the available data, it was not feasible to clearly resolve the low frequency component.
{The interpretation of the observations in terms of a ZF finds support from the observation of an enhancement of Long Range Correlations at the L--H transition~\cite{Hidalgo:2009} and predator-prey type oscillatory behaviour during the I phase~\cite{Estrada:2010c}.}  

{Taking the bicoherence as a measure of the ZF amplitude for simplicity, we make the following observations.
In the case of the L--H transition, the bicoherence appears well before the transition, reaches a maximum value close to the transition itself, and continues some time after the transition, although with gradually decreasing amplitude (Fig.~\ref{serie_101_42}).
This temporal sequence of events coincides qualitatively with the numerical results for the development of the ZF shown in Fig.~7d of \cite{Miki:2012}.
In addition, the simulated ZF in the cited work initially peaks in a narrow radial range, and then expands in radius through the time of the transition, similar to what is reported here for the bicoherence in Fig.~\ref{serie_101_42}.
In the case of the L--I transition, the simulated ZF appears after the L--I transition and expands radially (Fig.~4b of \cite{Miki:2012}), similar to the observations reported here for the bicoherence in Fig.~\ref{serie_100_35}.
Finally, we find that the bicoherence is initiated at a rather sharply defined perpendicular wavenumber ($k_\perp \simeq 10$ cm$^{-1}$), followed by an expansion in $k_\perp$ space, occurring immediately after the fast L--H transition, or some tens of milliseconds after the start of the I phase in the L--I transition.
}

Putting these results into context, we note that this is the first time wave number resolved bicoherence during L--H transitions is reported.
This important result indicates which spatial scales are important in the zonal flow coupling producing the L--H transition.

\section*{Acknowledgements}
The authors would like to express their gratitude for continued support by the TJ-II team.
Research sponsored in part by DGICYT of Spain under project Nr.~ENE2010-18409, and in part by the Ministerio de Econom\'ia y Competitividad of Spain under project Nr.~ENE2012-30832.


\section*{References}


\end{document}